\documentclass[twocolumn,amsmath,amssymb,pre]{revtex4}
\usepackage{graphicx,color,longtable}

\begin{document}

\makeatletter

\title{The role of quench rate in colloidal gels}

\author{C. Patrick Royall}
\affiliation{School of Chemistry, University of Bristol, Bristol, BS8
1TS, U.K}

\author{Alex Malins}
\affiliation{Bristol
Centre for Complexity Science, University of Bristol, Bristol, BS8
1TS, UK.}

\begin{abstract}
Interactions between colloidal particles have hitherto usually been fixed by the suspension composition. Recent experimental developments now enable the control of interactions \emph{in-situ}. Here we use Brownian dynamics simulations to investigate the effect of controlling interactions on gelation, by ``quenching'' the system from an equilibrium fluid to a gel. We find that, contrary to the normal case of an instantaneous quench, where the local structure of the gel is highly disordered, controlled quenching results in a gel with a higher degree of local order. Under sufficiently slow quenching, local crystallisation is found, which is strongly enhanced when a monodisperse system is used. The higher the degree of local order, the smaller the mean squared displacement, indicating an enhancement of gel stability.
\end{abstract}

\maketitle

\section{Introduction}

Gelation is among the most striking features of soft matter \cite{zaccarelli2007,poon2002,ramos2005sdg}.
Although many everyday materials are readily classified as gels, from
toothpastes to yoghurts, a deep understanding of the gel state
remains a challenge. In particular, three questions that might be
tackled are, (i) what distinguishes a gel from a glass, (ii) which
materials can form gels, (iii) for a given material, what are the
requirements for gelation? This work, which explores controlled quenching in colloidal gels with Brownian dynamics simulations is relevant to (iii).

For the first question, the identification of gelation with the crossing
of a liquid-gas spinodal \cite{lu2008,verhaegh1997,zaccarelli2007}
provides a working definition to distinguish gels from other dynamically
arrested states, namely glasses. For the second, it seems that an effective attraction is required. Systems with short-ranged attractive interactions
(up to around 10\% of the particle size) are well known to undergo
gelation \cite{zaccarelli2007,poon2002,ramos2005sdg}. In the case
of longer-ranged attractions in which the liquid state is stable,
shallow quenches below the gas-liquid spinodal lead to complete phase
separation \cite{testard2011,zhang2012}. Systems with short-ranged attractions
\emph{and} long-ranged
repulsions can also exhibit similar behaviour \cite{campbell2005,zaccarelli2007,klix2010,puertas2004}.
However the long-ranged repulsion significantly complicates the energy
landscape, leading for example to degenerate mesophases \cite{tarzia2006,tarzia2007,archer2007}.
Hereafter, we focus on systems without long-ranged repulsions, so the gels we consider
are explicitly thermodynamically metastable.

Almost all gel literature concerns soft materials, that is to say, multicomponent
systems of one or more mesoscopic component (colloids, nanoparticles,
polymers) suspended in a fluid. However, gel-forming systems are often treated
by integrating out the degrees of freedom of the smaller
components (solvent molecules, small ions, polymers, etc) and considering
an effective one-component system \cite{likos2001}. A reasonable
question then is - can true one-component (molecular) systems form
gels? Although to our knowledge no experiment has yet been performed,
$C_{60}$ exhibits a property associated with a good gel-former, namely
an attraction whose range is short compared to the molecular size. $C_{60}$
is predicted to form a gel \cite{royall2011c60}, and even the longer-ranged Lennard-Jones
model (relative to the molecular diameter) will form a gel under a sufficiently deep quench \cite{testard2011}.

\begin{figure}
\includegraphics[width=65mm]{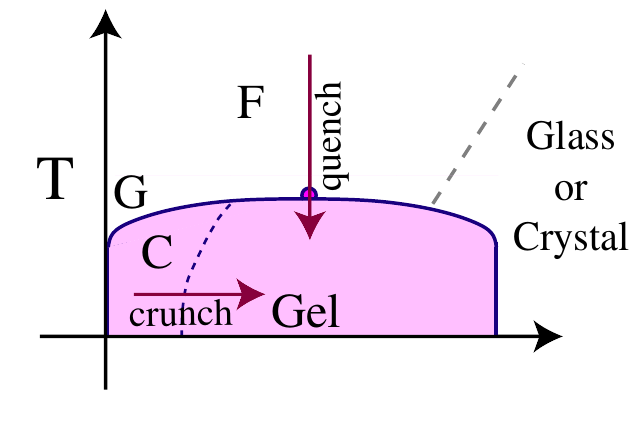}
\caption{\textbf{(colour online) Routes to gelation in short-ranged attractive systems}.
Gelation results upon quenching, provided the quench rate is sufficient
to avoid phase separation (shaded region). Crunching of isolated clusters
provides an addition route to form a gel where accessible quench rates
are too slow to prevent demixing. Here {}``C'' denotes isolated
clusters, ``G'' gas, ``F'' (supercritical) fluid.}
\label{figSchematic} 
\end{figure}

This brings us to the third question - the requirements for gelation
in a given material which forms the subject of this article. The gels we consider
are intrinsically non-equilibrium, and therefore, the means of preparation
is important. Here we shall consider a system with a relatively short-ranged
attraction, in which the equlibrium state is gas-crystal phase coexistence.
Forming a gel, therefore, means avoiding phase separation to a gas
and a crystal. Moreover, the resulting state must be dynamically
arrested, which also requires that it percolates \cite{cates2004}.
Possible routes to gelation are illustrated in Fig. \ref{figSchematic}.
The most obvious route is quenching, reducing the (effective) temperature.
Quenching must be carried out with sufficient speed that phase separation
does not occur. For soft matter systems, this is often straightforward. 

In colloids for example, where the absolute temperature
is typically held at 298 K, \emph{effective} temperature is varied
by changing the interaction strength. The effective temperature is
then related to the depth of the attractive well of the interaction
potential, and is often fixed for a given sample. Scanning a phase
diagram then requires preparation of a considerable number of different
samples, each with its own effective temperature. Following preparation of a sample, 
for example by shear, the system may be said to have undergone an (uncontrolled) instantaneous quench.  A typical example is a colloid-polymer
mixture, where the effective temperature is set by the polymer concentration
(through the depletion interaction) \cite{asakura1954}. In this case, the interactions are mediated by the polymer on much shorter timescales that the larger (and slower) colloids, and it is 
reasonable to consider an instantaneous quench as a route to gelation.

At smaller lengthscales, in nanoparticle and molecular systems, two main effects come into play. Firstly interactions are often reasonably constant over the temperature range of interest, but temperature is used as a control parameter. A consequence then is that quench rates \emph{cannot} be instantaneous is often the case for colloids. This is significant, as
 in the case of small nanoparticles
and molecular systems, the dynamics are often fast enough that quench
rates which avoid phase separation can be technically challenging
\cite{royall2011c60}. In this case, crunching, where a gas of clusters
is compressed, remains a route by which gels can be realised.

However, since gels are usually formed of soft matter with mescoscopic components, it is natural that most work on gelation assumes that the interactions between the particles were fixed. In other words, that quenches were instantaneous. This is typically the case in both experiments \cite{zaccarelli2007,ramos2005sdg}  and computer simulation \cite{zaccarelli2007,puertas2004,soga1998,darjuzon2003,fortini2008}.
Recent developments in controlling colloid-colloid interactions allow the interactions
(and thus the effective temperature) to be changed at will, even on timescales much faster
than the colloid dynamics, such that controllable quenches may be carried out \cite{assoud2009}. It has also become possible to control attractive interactions between colloids 
such as temperature-dependent depletion attractions \cite{alsayed2004,taylor2012}, multiaxial electric fields
\cite{elsner2009} and the critical Casimir effect \cite{hertlein2008,guo2008,bonn2009}. This
opens the way to consider the role of controlling effective temperature
(by changing interactions) in an experiment and thus enables us to consider
the role of quench rate in colloidal gelation. Varying the quench rate has been
carried out in gelation in molecular systems \cite{royall2011c60}. There,
using molecular dynamics simulations, a high quench rate was found
to be necessary to prevent phase separation to equilibrium gas-crystal
coexistence. Here we shall use Brownian dynamics simulations to model colloidal gelation in which we shall consider the effect of quench rate upon the gels formed.

This paper is organised as follows. First we introduce the model system,
which is a representation of a colloid-polymer mixture using Brownian
dynamics simulations \cite{royall2008g}. We then consider the response
of the system to a conventional treatment of an instantaneous quench
where the effective temperature is fixed at the outset. The system
is characterised through a novel analysis of local structure we have developed, the topological
cluster classification (TCC) \cite{royall2008g,williams2007}. The
effect of quench rate is then presented in both polydisperse and monodisperse systems. Finally we discuss the implications
of our findings.

\section{Methods}

\subsection{Model}

We model a colloid-polymer mixture with Brownian dynamics simulations.
In the experimental system upon which we base our simulations, residual electrostatic interactions are
screened~\cite{royall2007,royall2008g} and are neglected here. We found good
agreement with simulations using the Morse potential\cite{royall2008g}.
The Morse potential reads

\begin{equation}
\beta u(r)=\beta\varepsilon e^{ -\rho_{0} \left( \sigma-r \right) }  \left(  e^{ -\rho_{0}(\sigma-r)} -2\right)
\label{eqMorse}
\end{equation}

\noindent where $\beta=1/k_{B}T$ the thermal energy. The potential
is truncated and shifted at $r=1.4\sigma$. The effective temperature
is controlled by varying the well depth of the potential $\varepsilon$. We set the range parameter
$\rho_{0}=25.0$. Colloidal particles are typically polydisperse,
and here we treat polydispersity by scaling $r$ in Eq. \ref{eqMorse}
by a Gaussian distribution in $\sigma$ with 4\% standard deviation
(the same value as the size polydispersity in the experimental system). In addition 
we consider the case of a monodisperse system.
We express time in units of the Brownian time $\tau_{B}=(\sigma/2)^{2}/6D$
which is the time taken for a particle to diffuse its own radius. Here
$D$ is the diffusion constant. In the simulations, $\tau_{B}\approx894$
time units. The time step is 0.03 simulation time units. Where the
effective temperature is fixed throughout the simulation (Section
\ref{sub:Instantaneous-quenches}), the runs are equilibrated for
$5.0\times10^6$ steps and run for a further $5.0\times10^6$ steps. These simulation runs
therefore correspond to approximately $168$ Brownian times or around
10 min, a timescale certainly comparable to experimental work. Except in the case of long quench times ($1.01\times10^{4}$ $\tau_B$) where 2048 particles were used, the system size was 10000 particles. We fix the packing fraction $\phi=\pi\rho\sigma^{3}/6=0.1$
throughout. At this packing fraction, quenching below the critical
point, which is approximately located at ($\beta\varepsilon^{c}\approx2.69$)
\cite{noro2000}, leads to a percolating network - i.e. a gel. The
critical point gives a reasonable estimate for gelation, since the
spinodal line is found to be very flat for short-ranged systems \cite{eliott1999}.
In the case of simulations where a quench is carried out, $\beta\varepsilon$ is increased 
linearly from unity to a chosen value, at which point the system is run for $5\times10^6$ time steps and analysed.

\subsection{Structural Analysis - Topological Cluster Classification}

\begin{figure}
\includegraphics[width=45mm]{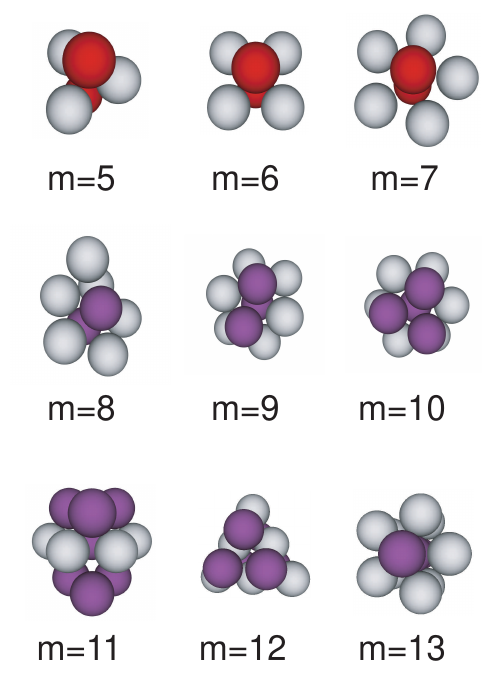}
\caption{\textbf{Clusters detected by the topological cluster classification.} 
These structures are minimum energy clusters of the Morse potential with $\rho_0=25.0$
\cite{doye1995}.}
\label{figTCC} 
\end{figure}

To analyse the structure, we identify the bond network using the Voronoi
construction with a maximum bond length of $1.4$ $\sigma$. Having identified the bond network, we use the Topological
Cluster Classification (TCC) to determine the nature of the cluster~\cite{williams2007,malins2012}.
This analysis identifies all the shortest path three, four and five
membered rings in the bond network. We use the TCC to find clusters
which are global energy minima of the Morse potential for the range we consider $(\rho_{0}=25.0)$, as listed in
~\cite{doye1995} and illustrated in Fig. \ref{figTCC}. 
In addition we identify the thirteen particle structures which correspond to 
FCC and HCP in terms of a central particle and its twelve nearest neighbours.
For more details see \cite{williams2007,malins2012}.

\section{Results}

\begin{figure*}
\includegraphics[width=140mm]{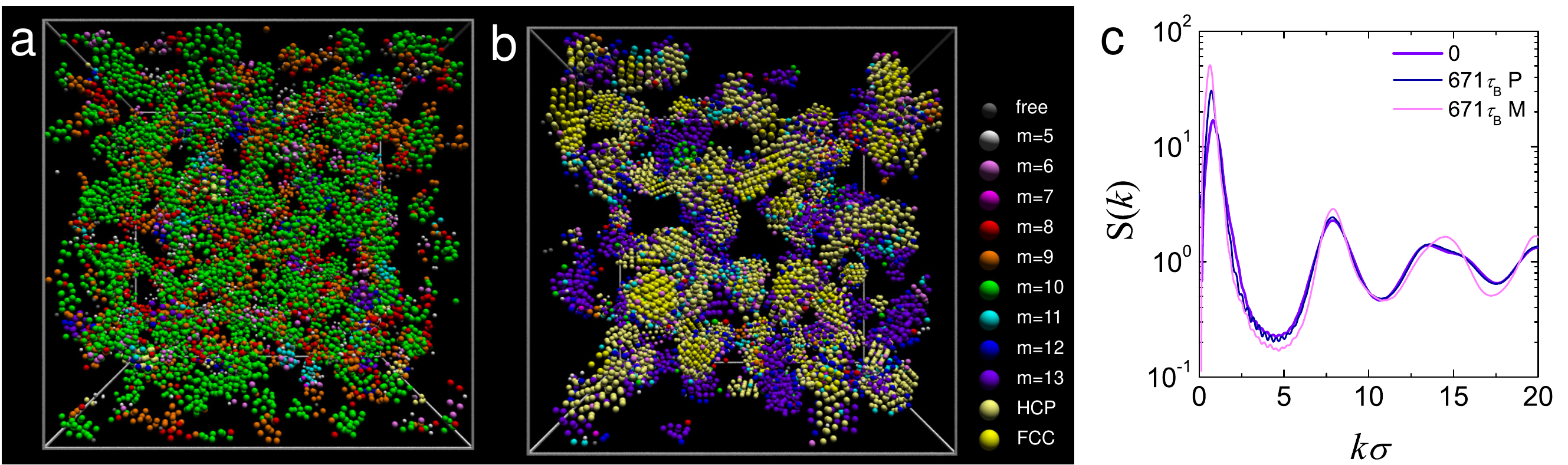}
\caption{\textbf{Simulation snapshots labelled by the topological cluster classification.} (a) Polydisperse system instantaneously quenched. (b) Monodisperse system with a quench time of 671 $\tau_B$. Both (a) and (b) correspond to an identical state point with $\beta \varepsilon = 5.0$. Color of particles on right of (b) denotes cluster in which the particle is identified by the TCC as shown in Fig. \ref{figTCC}. Note the vastly increased population of particles in crystalline environments in the case of the monodisperse system with a quench time of 671 $\tau_B$.
(c) Static structure factor $S(k)$ for state points shown in (a) (purple line) and (b) (pink line) along with a polydisperse system with a quench time of  671 $\tau_B$ (navy line).}
\label{figPretty} 
\end{figure*}

The results section is organised as follows. We begin by presenting
a brief overview of the effect of quenching. We consider the 
conventional case of instantaneous quenching, and provide an analysis
from the perspective of local structure. We next investigate a finite quench rate, and the effect of varying quench rate on
the resulting structure and dynamics of the gel. Having established the general nature of 
quenching on the local structure, we examine the special case of a monodisperse system.
Dynamic data is then
employed to investigate relative stabilities of the two routes to
gelation - instantaneous quenching and quenching at finite rates.

An overview of the system is shown in Fig. \ref{figPretty}.  These snapshots show the network structure that is developed by colloidal gels at a packing fraction $\phi=0.1$ and attraction $\beta \varepsilon = 5.0$. While both are networks, the detail of the local structure is markedly different, as revealed by the topological cluster classification. The instantaneously quenched (polydisperse) system [Fig. \ref{figPretty}(a)] is dominated by a 10 membered fivefold symmetric local structure, while the monodisperse system with a quench time 671 $\tau_B$ shows a significant degree of local crystallisation but still retains a similar overall network structure. A slight degree of coarsening is apparent in the static structure factor $S(k)$ in Fig. \ref{figPretty}(c) relative to the instantaneous quench. We also note that, for the same quench time, a monodisperse system appears to coarsen somewhat more than a polydisperse system. However, the change in $S(k)$ appears much less dramatic than that in local structure illustrated in Fig. \ref{figPretty}(a) and (b).

\subsection{Instantaneous quenches in polydisperse systems - a local structural analysis}

\label{sub:Instantaneous-quenches}

\begin{figure*}
\includegraphics[width=160mm]{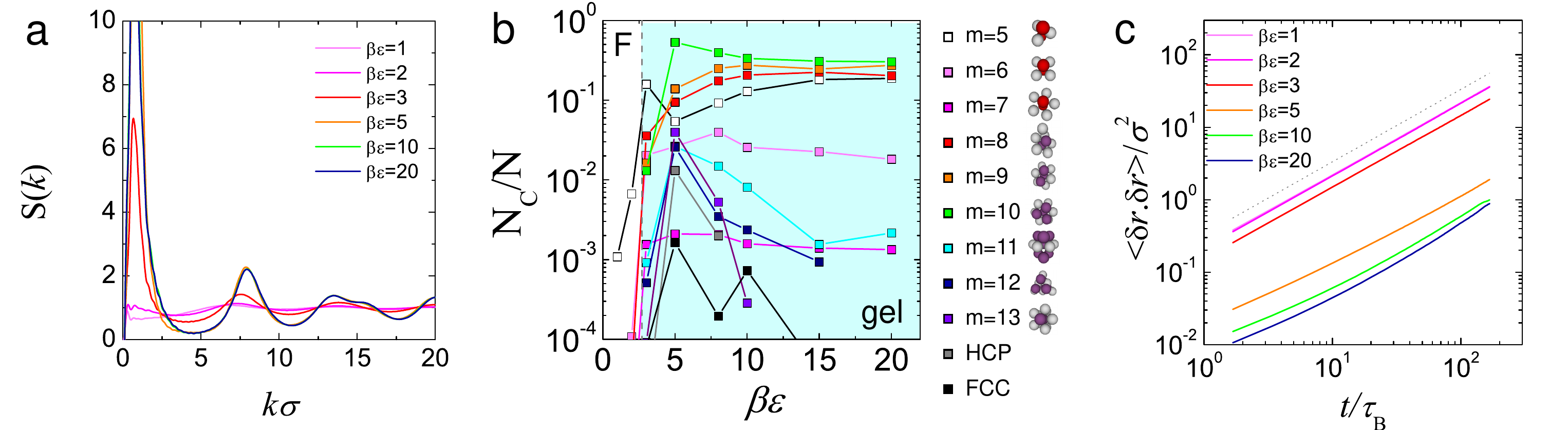}
\caption{\textbf{The conventional situation: instantaneous
quench.} (a) The static structure factor $(S(k))$ for varying strengths
of attraction. Where gelation occurs on the simulation timescale ($\beta\varepsilon>3$),
the $S(k)$ are indistinguishable for practical purposes. (b) The
topological cluster classification reveals changes in local structure
upon quenching. Dashed line denotes the strength of attraction at criticality $\varepsilon^c$ which approximately locates gelation. $N_c$ is the number of particles in a given cluster, $N$ is the total number of particles.
 (c) mean squared displacement for the well depth $\beta\varepsilon$
(inverse effective temperature) indicated. Dashed grey line has a
slope of unity (diffusive motion).}
\label{figInstant} 
\end{figure*}

We now examine instantaneous quenches in more detail.
A conventional structural analysis is presented in Fig. \ref{figInstant}(a)
in the form of the static structure factor $S(k)$. This illustrates
the typical behaviour of systems undergoing dynamical arrest: 
the structure in the fluid ($\beta\varepsilon\leq3$) responds to changes in
effective temperature, in that the low $k$ region of $S(k)$ increases
strongly as phase separation is approached. Conversely, upon dynamical arrest for 
$\beta \varepsilon \ge 5$, no further change in the static structure is seen. Thus one might
conclude that under instantaneous quenching, all gels with $\phi=0.1$ have the same local structure. 

In fact this is not so, as shown in Fig. \ref{figInstant}(b). Here
the topological cluster classification is shown for the same data.
This analysis reveals the generic nature of gelation at the single-particle
level: at high effective temperatures, in the ergodic fluid phase,
few clusters are found. Those few that are present are predominantly
five-membered triangular bipyramids. Upon gelation, the cluster population
rises very sharply. This indicates that the particles condense into
clusters which comprise the gel network [Fig. \ref{figPretty}(a) and (b)]. However it is the type
of the clusters which reveal the local structure of the gel. Although the
equilibrium state is gas-crystal coexistence, even at the level of
the few particles associated with a cluster, the kinetic 
pathway to crystallisation is arrested
and very few particles ae found in a local HCP or FCC environment. 
Upon deeper quenching, the system
moves further from equilbrium, and even fewer particles in local
crystalline environments are found. Rather than crystalline environments,
particles are found in clusters of between 8 and 10 particles. As illustrated
in Fig. \ref{figTCC}, these are built around pentagonal bipyramids
and are thus five-fold symmetric. Such five-fold symmetry has long
been believed to suppress crystallisation \cite{frank1952}. We note
that these same clusters are prevalent in dense fluids of both Morse
particles and hard spheres \cite{taffs2010jcp}. This feature of
crystallisation suppressed by five-fold symmetric clusters is found
also in experiments on colloidal gels \cite{royall2008g}, simulations
of molecular gels \cite{royall2011c60}, and isolated clusters \cite{malins2009},
where in the case of the latter, the ground state is one of the clusters
considered by the TCC. However, suppression of crystallisation is not universal, experiments in which locally crystalline clusters are found have also been reported \cite{zhang2011cluster}. 

The change in dynamics upon gelation for instantaneous quenches is
shown in the mean squared displacement (MSD) data in Fig. \ref{figInstant}(c).
Two observations are important concerning this plot. The first is
that all state points are diffusive. There is little indication
of a plateau that one might associate with slow dynamics. The second
is that although in the fluid state, the system is rather insensitive
to the effective temperature with only a slight decrease in MSD upon
increasing the strength of the attraction. Upon gelation there is
a sudden drop in mobility of more than an order of magnitude. Although
MSD indicates diffusive motion, the magnitude of the motion is small,
only reaching the particle lengthscale after around 100$\tau_{B}$.
As significant as this drop in mobility is, some restructuring cannot
be ruled out, indicating that these gels age.

\subsection{Finite quenches in polydisperse systems - structure}

\begin{figure*}
\includegraphics[width=160mm]{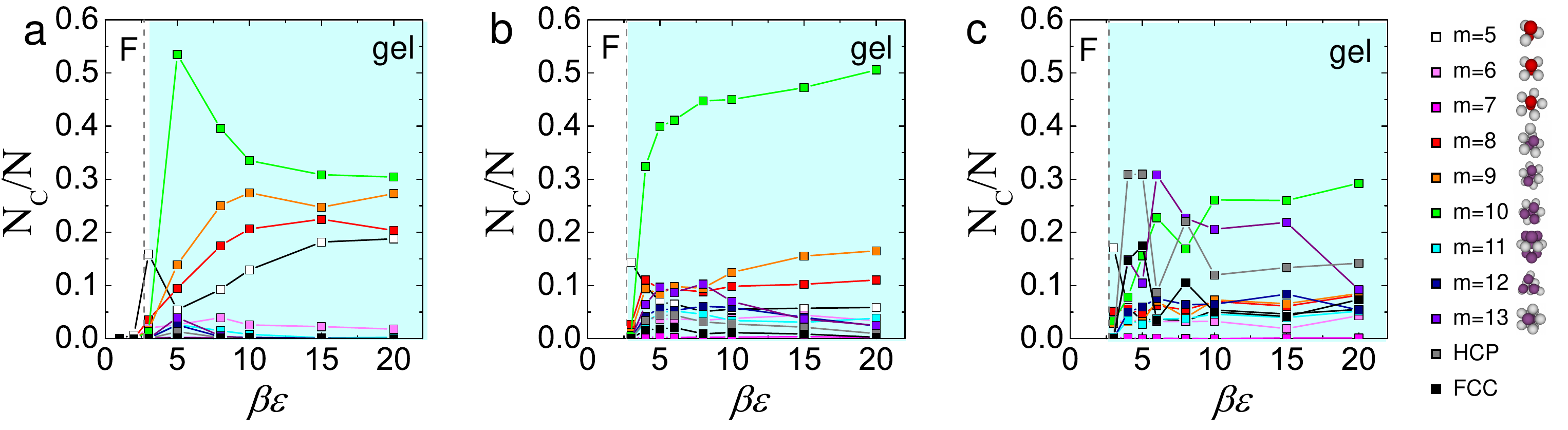}
\caption{\textbf{The effect of a quench rates on the local
structure in polydisperse systems} (a) TCC analysis for a quench time of 671$\tau_{B}$. (b)
TCC analysis for a quench time of $1.01\times10^{4}$$\tau_{B}$.
(c) Data from an instantaneous quench. Dashed lines denote $\varepsilon^c$}
\label{figQuench} 
\end{figure*}

Let us consider finite quench rates. We shall use our local
structural analysis, the TCC. As might be surmised from Fig. \ref{figInstant}(a),
the static structure factor is not very sensitive to quench rate.
TCC analyses of data for quench times of 671$\tau_{B}$ and $1.01\times10^{4}$$\tau_{B}$
are shown in Fig. \ref{figQuench}(a) and (b) respectively. In the corresponding experimental system, the quench times are around 30 minutes and 8 hours respectively.
For comparison, instantaneous quench data is plotted in Fig. \ref{figQuench}
(c). Here we quench from a fluid at $\beta\epsilon=1$ to the final
value of $\beta\epsilon$ as shown in Fig. \ref{figQuench}(a) and
(b). Each effective temperature therefore corresponds to a different
simulation run. We begin by considering the moderate quench time of
671$\tau_{B}$. The generic behaviour is similar to that of an instantaneous
quench {[}Fig. \ref{figQuench}(c)], with a sharp rise in cluster
population upon gelation (the very small cluster population in the
equilibrium fluid $\beta\varepsilon \lesssim\beta\varepsilon^{c}$ is
of course unaffected by the quench protocol). For shallow quenches
($\beta\varepsilon\sim5$), there is a qualitative similarity: the
structure is dominated by the 10-membered cluster. However, upon
quenching deeper a significant difference emerges. The effect is
considerable and is summarised as follows: In the case of a moderate
quench time, the fivefold symmetric 10-membered cluster population
shows a rapid rise upon gelation up to a population of around $N_{c}/N\approx0.42$
and continues to increase slowly upon deeper quenching, finally reaching
a value of $N_{c}/N\approx0.5$ for $\beta\epsilon=20$. Very different
behaviour is seen in the case of the instantaneous quench. Although
the 10-membered cluster again exhibits strong rise upon gelation,
its population \emph{falls} upon deeper quenching, finally reaching
a value around $N_{c}/N\approx0.3$ for $\beta\epsilon=20$. Other
smaller clusters, notably eight- and nine-membered clusters form a
very significant fraction of the population in the case of deep quenches.

For a long quench time ($1.01\times10^{4}$ $\tau_{B}$), Fig. \ref{figQuench}(b)
shows a further change in local structure. Higher-order clusters again
are favoured. In particular, for $4\leq\beta\varepsilon\leq5$, the
most popular cluster is the HCP crystalline environment, with a value
of $N_{c}/N\approx0.3$. In other words, around 30\% of the system
has succeeded in reaching, or at least getting close to, local equilibrium.
Even at deeper quenches, although the crystal population is smaller
($N_{c}/N\approx0.15$), much more of the system has crystallised
compared to shorter quench times. The rise in HCP environments is
accompanied by a rise in the 13-membered five-fold symmetric cluster.
This pentagonal prism which is the minimum energy structure for 13
Morse particles is similar to - but distinct from - the icosahedron.

\subsection{Finite quenches in monodisperse systems - structure}

\begin{figure*}
\includegraphics[width=160mm]{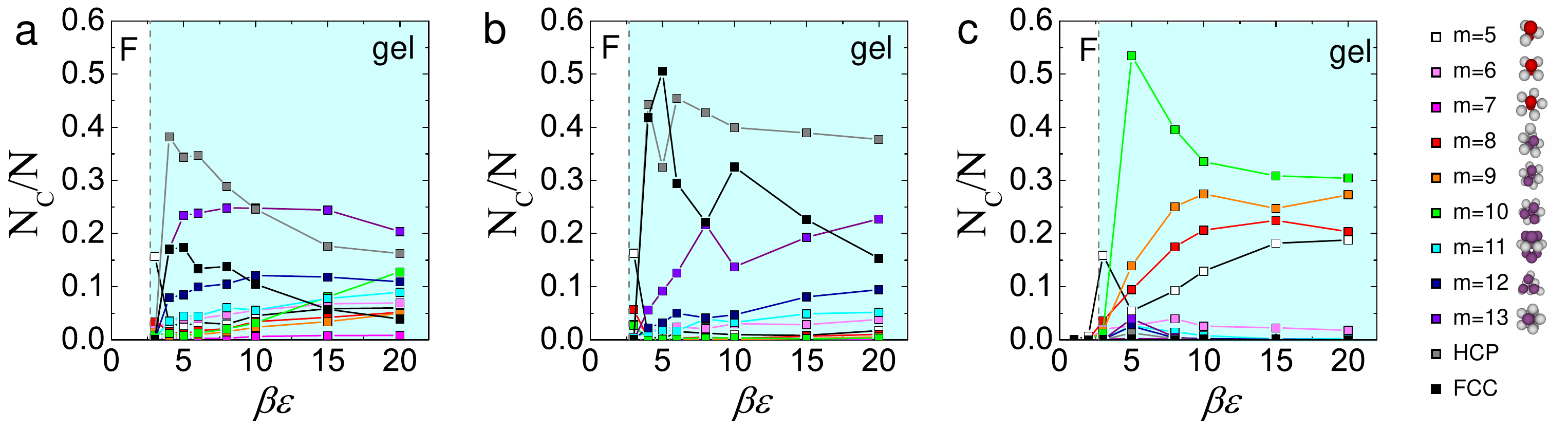}
\caption{ \textbf{The effect of a quench rates on the local
structure in monodisperse systems} (a) TCC analysis for a quench time of 671$\tau_{B}$. (b)
TCC analysis for a quench time of $1.01\times10^{4}$$\tau_{B}$.
(c) Data from an instantaneous quench. Dashed lines denote $\varepsilon^c$}
\label{figQuenchM} 
\end{figure*}

Polydispersity is known to suppress crystallisation. We therefore consider the effect of setting the polydispersity to zero (a monodisperse system), keeping all other conditions the same. The effect on the local structure is shown in Fig. \ref{figQuenchM}. Except that we now consider a monodisperse system, the situation is  identical to Fig. \ref{figQuench}. Perhaps the most striking result is the instantaneous quench, shown in Fig. \ref{figQuenchM}(c). The similarity to the polydisperse case is remarkable, and crystallisation is strongly suppressed. This is consistent with the idea that under instantaneous quenching, very little re-arrangement is possible, particles remain kinetically trapped in the clusters they form upon condensation, and lack the thermal energy to rearrange. 

The case of the finite quenches [$671$ $\tau_B$ and $1.01\times10^{4}$ $\tau_B$ in Fig. \ref{figQuenchM}(a) and (b) respectively] is very different indeed. Unlike the dominance of the five-fold symmetric 10-membered cluster in the polydisperse case, here the crystalline HCP and FCC clusters are the most abundant. Like the polydisperse system, the degree of crystallisation appears to be increased in the case of slow quenching ($1.01\times10^{4}$ $\tau_B$). As shown in Fig. \ref{figPretty}(b), this monodisperse system can largely reach local equilibrium, however it retains a network structure: a crystalline gel.

\subsection{Finite quenches - dynamics}

\begin{figure}
\includegraphics[width=85mm]{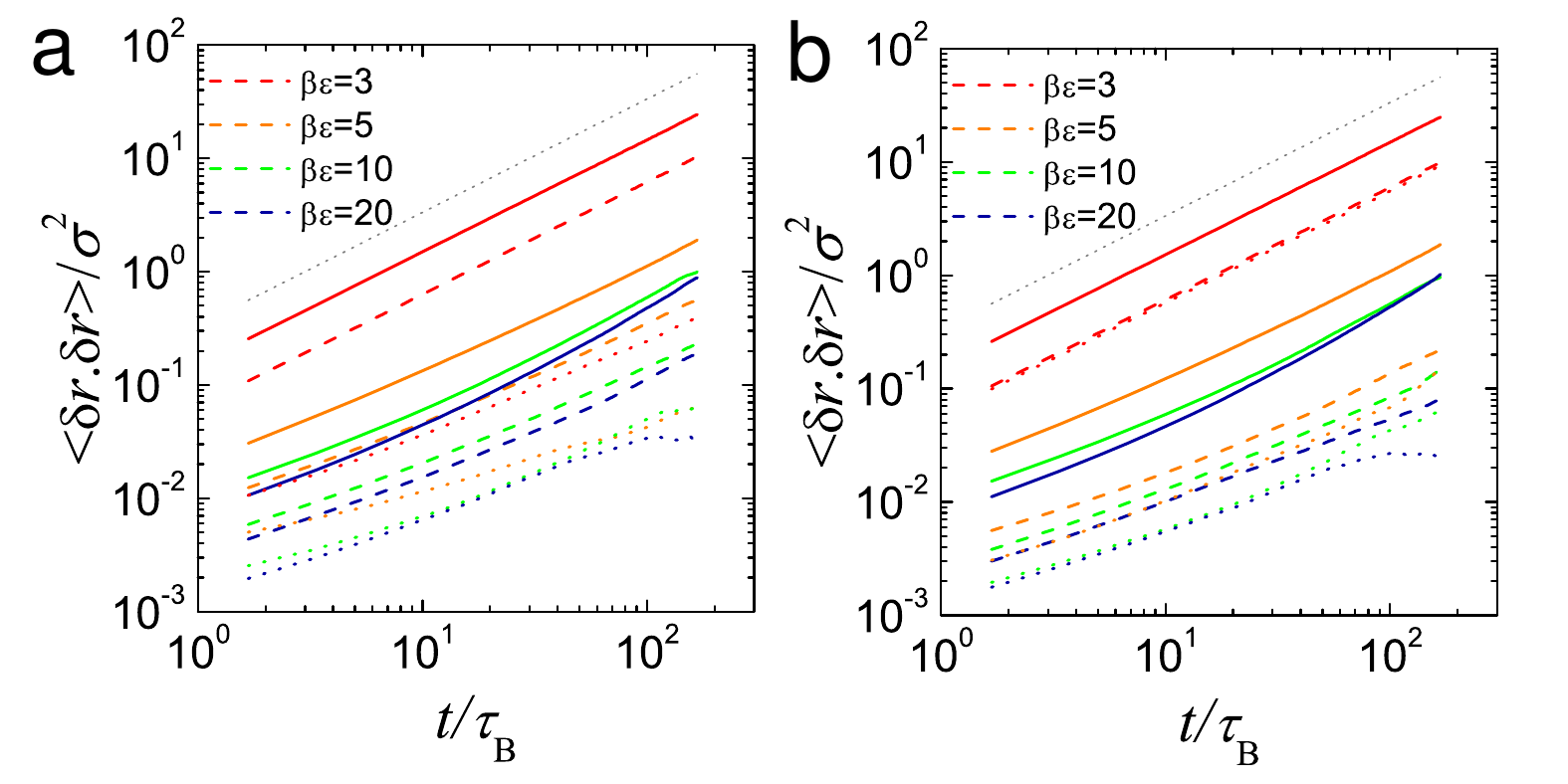}
\caption{ \textbf{The effect of a quench times on the} \textbf{dynamics.}
(a) Polydisperse system (b) monodisperse system.
Mean squared displacement for quench depth and quench times. Solid
lines - instantaneous quench, dotted lines, moderate (671$\tau_{B}$)
quench time, dashed lines, long ($1.01\times10^{4}$$\tau_{B}$) quench
time.}
\label{figMSDQuench} 
\end{figure}

The effect of quench times upon the dynamcs is shown in Fig. \ref{figMSDQuench}.
As in the instantaneous case, the mean squared displacement appears diffusive,
with no indication of a plateau. However, as noted above, the magnitude
of the MSD is small, with the exception of $\beta\varepsilon=3$.
For the finite quench times, the MSD drops significantly compared to
the instantaneous quench time. For the long quench time, ($1.01\times10^{4}$$\tau_{B}$),
this drop is very considerable indeed, with a reduction of more than
an order of magnitude. At the very least, this suggests that the finite
quench time leads to a gel with a much higher stability. Similar effects are found in the case of monodispersity, which seems consistent with the increase in crystalline order observed above.

\section{Discussion}

We have shown that colloidal gels quenched slowly exhibit local structural differences compared to the same system quenched instantaneously (in other words, where the particle interactions are fixed). The local structural motifs of instantaneously quenched gels are are five-fold symmetric, and are also found in dense equilibrium fluids for this system \cite{taffs2010jcp}. It seems reasonable to suppose that, upon quenching, the system begins to phase separate, and the dense fluid phase - whose arrest drives gelation - is structurally similar to high-density fluids. In particular, the local structure is dominated by fivefold symmetric clusters whose principle motif is a five-membered ring ($7 \leq m \leq 10$ in Fig. \ref{figTCC}). Perhaps surprisingly, the local structure upon instantaneous quenching is rather insensitive to polydispersity.

\subsection{Structure and finite quench rates}

Quenching at a finite rate on the other hand gives the system some chance to reach thermodynamic equilibrium as it is ``cooled''. Specifically, in equilibrium, phase separation is expected for $\varepsilon > \varepsilon^c$. As mentioned in the introduction, here we are concerned with a system with short range attractive interactions in which the gel is explicitly metastable to crystal-gas coexistence.
However, even for a quench rate of $4.0\times10^{-4}$ $\beta \varepsilon \tau_B ^{-1}$, no such phase separation is observed, although some coarsening is observed for a moderate quench time of $671$ $\tau_B$. In the case of experiments on colloid-polymer mixtures, due to the change in polymer size with temperature $1^\circ$C corresponds to a change in attraction of order $k_BT$. The resulting change in effective temperature (due to the polymer-induced depletion attraction) corresponds to a quench rate of just $\sim 10^{-3}$ $^\circ$C$s^{-1}$. Thus the overall feature of colloidal gelation is remarkably robust to temperature quenching. Or, equivalently, colloidal model systems are very slow to equilibrate.

Although the system continues to gel, even with moderate quench rates (quench time $671$ $\tau_B$), the local structure is a strong function of cooling rate. In particular, the structure of the gel is rather constant with quench depth [Fig. \ref{figQuench}(a)]. In other words, in the early stages of gelation, soon after the system has become metastable and begin to phase separate, the particles assemble into 10-membered clusters. Upon deeper quenching, this local structure is preserved, as the particles gradually lose their mobility. By contrast, systems subjected to instantaneous quenches are less able to self-organise into such large clusters and, particularly at deeper quenches, a substantial number of smaller clusters are present. Of particular note is the 5-membered triangular bipyramid. This is formed from two tetrahedra, and the tetrahedron is the simplest rigid assembly of spheres in 3D. In other words, instantaneous quenches lead to very simple structures, which suggest rapid formation with no further re-arrangement, such as is the case for diffusion-limited cluster aggregation.

The opposite case then, is a slow quench ($1.01\times10^{4}$ $\tau_B$). Here, although full phase separation does not occur, a very considerable number of particles are able to reach local equilibrium and crystallise. We have based our discussion on a slightly polydisperse system as found in experiments \cite{royall2008g, royall2007}. This has two main consequences. Firstly, crystallisation should be suppressed. Although hard spheres with a polydispersity of 4\% as we have employed here do crystallise, short-ranged attractive systems are known to be more sensitive to polydispersity \cite{poon2002}. Secondly, the relationship between polydispersity and the eventual approach to equilibrium is complex. What is known is that polydisperse hard spheres phase separate into less polydisperse daughter populations \cite{wilding2011}, and it seems reasonable to suppose that similar behaviour might occur here. At the very least, this suggests that full equilibration would take a long time.  Conversely, in the monodisperse case, a far higher degree of crystallisation is found, even though for instantaneous quenches the local structure seems insensitive to polydispersity. However we note that previous work did indeed find some crystallisation in Brownian dynamics simulations of monodisperse attractive particles where quenching was instantaneous, but this tended to be supressed by gelation - crystallisation occured at weaker attractions than $\varepsilon^c$ \cite{fortini2008}. Such observations are consistent with measures of the self-assembly from a perspective of local reversibility as determined by deviations from the fluctuation-dissipation theorem \cite{klotsa2011}, where optimal assembly (crystallisation) occurs at $\varepsilon<\varepsilon^c$.

\subsection{Dynamics and finite quench rates}

The drop in MSD induced in the same system by varying the quench rate is remarkable, and surprising in its extent. At one level, it underlines the need to use finer probes of structure than $S(k)$ in order to unravel the origins of the dynamical behaviour. We can rationalise the reduction in MSD as consistent with the system being better equilibrated, i.e. lower in its energy landscape. As intuitive as this observation is, backed up by structural observations consistent with improved equilibration, the strong dependence of the dynamics is nonetheless worthy of further investigation. Moreover, the MSD data raises additional questions.

For example, why are the MSDs diffusive? Although the sampling times are quite short, the absence of any plateau can be interpreted as indicative of further irreversible structural rearrangement. However, that all the MSDs have such similar slopes almost regardless of quench depth and time remains a curiousity.

A further comment on the approach to equilibrium and stability concerns aging. While this has already recieved attention for instantaneuous quenching \cite{darjuzon2003,foffi2005jcp}, an important question concerns the link between aging and quench rate. Certainly, instantaneously quenched systems are higher in the free energy landscape and exhibit more mobility. Longer quench times produce systems lower in the energy landscape with less mobility, i.e. precisely the consequences of ageing \cite{ramos2005sdg}. Can quenching be related to aging? Is it directlty equivalent? Connecting our measures of local structure with dynamical quantities such as MSD provides a means to tackle these questions.

\subsection{The role of dynamics}

Before closing, a few words on the role of the dynamics employed are in order. In short, this is the observation that, while here we have used Brownian dynamics and seen no phase separation,  in a comparable system with molecular dynamics (MD) \cite{royall2011c60}, \emph{it was hard to prevent complete phase separation}. While the systems were not accurately mapped to one another, (the MD system has a small region where the liquid is thermodynamically stable), the extent of the difference in behaviour seems to warrant further investigation, although some previous work did not reveal a very large difference between the two forms of dynamics \cite{foffi2005jcp}. Can a system with Newtonian dynamics ``know'' its way through the energy landscape so much better than a system with Langevin dynamics that it is hard to kinetically trap the former and apparently impossible to avoid kinetic trapping in the latter? We hope to answer this question in the near future, and also to develop the results we have presented here with larger scale simulations that can explore the role of finite size effects. Finally, we note that solvent-mediated hydrodynamic interactions have been shown to have a profound effect upon gelation and may also influence local structure \cite{furukawa2010}.

\section{Conclusions}

We have carried out Brownian dynamics computer simulations of a colloidal gel. By ``cooling'' the system from the stable fluid to a metastable gel, we have explored the role of quench rate in colloidal gels. Our results indicate that, despite its metastable nature, gelation is surprisingly robust. All quench rates we have been able to perform resulted in a gel, none exhibited phase separation. Conversely, the local structure accessed through the topological cluster classification is a strong function of quench rate. Decreasing the quench rate enables the system to access larger structures associated with lower basins in the energy landscape, and, for the slowest rates we access, a considerable part of the system, can reach local equilibrium and crystallise. Monodisperse systems exhibit a much higher degree of crystallisation than only slightly polydisperse systems. This is reflected in the dynamics: the mean squared displacement drops markedly \emph{for the same state point} when a slow quench is employed. We therefore conclude that local structure, as measured by the TCC, is strongly coupled to the dynamics.
Our results suggest that controlled quenching, as is beginning to become possible in experiments on colloidal model systems, may be a means by which colloidal gels can be stablised, and products based on gels may enjoy an extended shelf life.

\section*{Acknowledgements}

We gratefully acknowledge stimulating discussions with Jens Eggers, Rob Jack, Hajime Tanaka and Stephen Williams.
A.M. is funded by EPSRC grant code EP/E501214/1. C.P.R. thanks the
Royal Society for funding. This work was carried out using the computational
facilities of the Advanced Computing Research Centre, University of
Bristol.


\begin{thebibliography}{42}
\expandafter\ifx\csname natexlab\endcsname\relax\def\natexlab#1{#1}\fi
\expandafter\ifx\csname bibnamefont\endcsname\relax
  \def\bibnamefont#1{#1}\fi
\expandafter\ifx\csname bibfnamefont\endcsname\relax
  \def\bibfnamefont#1{#1}\fi
\expandafter\ifx\csname citenamefont\endcsname\relax
  \def\citenamefont#1{#1}\fi
\expandafter\ifx\csname url\endcsname\relax
  \def\url#1{\texttt{#1}}\fi
\expandafter\ifx\csname urlprefix\endcsname\relax\def\urlprefix{URL }\fi
\providecommand{\bibinfo}[2]{#2}
\providecommand{\eprint}[2][]{\url{#2}}

\bibitem[{\citenamefont{Zaccarelli}(2007)}]{zaccarelli2007}
\bibinfo{author}{\bibfnamefont{E.}~\bibnamefont{Zaccarelli}},
  \bibinfo{journal}{J. Phys.: Condens. Matter} \textbf{\bibinfo{volume}{19}},
  \bibinfo{pages}{323101} (\bibinfo{year}{2007}).

\bibitem[{\citenamefont{Poon}(2002)}]{poon2002}
\bibinfo{author}{\bibfnamefont{W.~C.~K.} \bibnamefont{Poon}},
  \bibinfo{journal}{Journal of Physics, Condensed Matter}
  \textbf{\bibinfo{volume}{14}}, \bibinfo{pages}{R859} (\bibinfo{year}{2002}).

\bibitem[{\citenamefont{Ramos and Cipelletti}(2005)}]{ramos2005sdg}
\bibinfo{author}{\bibfnamefont{L.}~\bibnamefont{Ramos}} \bibnamefont{and}
  \bibinfo{author}{\bibfnamefont{L.}~\bibnamefont{Cipelletti}},
  \bibinfo{journal}{Journal of Physics: Condensed Matter}
  \textbf{\bibinfo{volume}{17}}, \bibinfo{pages}{R253} (\bibinfo{year}{2005}).

\bibitem[{\citenamefont{Lu et~al.}(2008)\citenamefont{Lu, Zaccarelli, Ciulla,
  Schofield, Sciortino, and Weitz}}]{lu2008}
\bibinfo{author}{\bibfnamefont{P.~J.} \bibnamefont{Lu}},
  \bibinfo{author}{\bibfnamefont{E.}~\bibnamefont{Zaccarelli}},
  \bibinfo{author}{\bibfnamefont{F.}~\bibnamefont{Ciulla}},
  \bibinfo{author}{\bibfnamefont{A.~B.} \bibnamefont{Schofield}},
  \bibinfo{author}{\bibfnamefont{F.}~\bibnamefont{Sciortino}},
  \bibnamefont{and} \bibinfo{author}{\bibfnamefont{D.~A.} \bibnamefont{Weitz}},
  \bibinfo{journal}{Nature} \textbf{\bibinfo{volume}{435}},
  \bibinfo{pages}{499} (\bibinfo{year}{2008}).

\bibitem[{\citenamefont{Verhaegh et~al.}(1997)\citenamefont{Verhaegh, Asnaghi,
  Lekkerkerker, Giglio, and Cipelletti}}]{verhaegh1997}
\bibinfo{author}{\bibfnamefont{N.~A.~M.} \bibnamefont{Verhaegh}},
  \bibinfo{author}{\bibfnamefont{D.}~\bibnamefont{Asnaghi}},
  \bibinfo{author}{\bibfnamefont{H.~N.~W.} \bibnamefont{Lekkerkerker}},
  \bibinfo{author}{\bibfnamefont{M.}~\bibnamefont{Giglio}}, \bibnamefont{and}
  \bibinfo{author}{\bibfnamefont{L.}~\bibnamefont{Cipelletti}},
  \bibinfo{journal}{Physica A} \textbf{\bibinfo{volume}{242}},
  \bibinfo{pages}{104} (\bibinfo{year}{1997}).

\bibitem[{\citenamefont{Testard et~al.}(2011)\citenamefont{Testard, Berthier,
  and Kob}}]{testard2011}
\bibinfo{author}{\bibfnamefont{V.}~\bibnamefont{Testard}},
  \bibinfo{author}{\bibfnamefont{L.}~\bibnamefont{Berthier}}, \bibnamefont{and}
  \bibinfo{author}{\bibfnamefont{W.}~\bibnamefont{Kob}},
  \bibinfo{journal}{Phys. Rev. Lett.}  (\bibinfo{year}{2011}).

\bibitem[{\citenamefont{Zhang et~al.}(2012{\natexlab{a}})\citenamefont{Zhang,
  Faers, Royall, and Bartlett}}]{zhang2012}
\bibinfo{author}{\bibfnamefont{I.}~\bibnamefont{Zhang}},
  \bibinfo{author}{\bibfnamefont{M.}~\bibnamefont{Faers}},
  \bibinfo{author}{\bibfnamefont{C.~P.} \bibnamefont{Royall}},
  \bibnamefont{and} \bibinfo{author}{\bibfnamefont{P.}~\bibnamefont{Bartlett}},
  \bibinfo{journal}{Submitted}  (\bibinfo{year}{2012}{\natexlab{a}}).

\bibitem[{\citenamefont{Campbell et~al.}(2005)\citenamefont{Campbell, Anderson,
  van Duijneveldt, and Bartlett}}]{campbell2005}
\bibinfo{author}{\bibfnamefont{A.~I.} \bibnamefont{Campbell}},
  \bibinfo{author}{\bibfnamefont{V.~J.} \bibnamefont{Anderson}},
  \bibinfo{author}{\bibfnamefont{J.~S.} \bibnamefont{van Duijneveldt}},
  \bibnamefont{and} \bibinfo{author}{\bibfnamefont{P.}~\bibnamefont{Bartlett}},
  \bibinfo{journal}{Phys. Rev. Lett.} \textbf{\bibinfo{volume}{94}},
  \bibinfo{pages}{208301} (\bibinfo{year}{2005}).

\bibitem[{\citenamefont{Klix et~al.}(2010)\citenamefont{Klix, Royall, and
  Tanaka}}]{klix2010}
\bibinfo{author}{\bibfnamefont{C.~L.} \bibnamefont{Klix}},
  \bibinfo{author}{\bibfnamefont{C.~P.} \bibnamefont{Royall}},
  \bibnamefont{and} \bibinfo{author}{\bibfnamefont{H.}~\bibnamefont{Tanaka}},
  \bibinfo{journal}{Phys. Rev. Lett.} \textbf{\bibinfo{volume}{104}},
  \bibinfo{pages}{165702} (\bibinfo{year}{2010}).

\bibitem[{\citenamefont{Puertas et~al.}(2004)\citenamefont{Puertas, Fuchs, and
  Cteas}}]{puertas2004}
\bibinfo{author}{\bibfnamefont{A.}~\bibnamefont{Puertas}},
  \bibinfo{author}{\bibfnamefont{M.}~\bibnamefont{Fuchs}}, \bibnamefont{and}
  \bibinfo{author}{\bibfnamefont{M.}~\bibnamefont{Cteas}}, \bibinfo{journal}{J.
  Chem. Phys.} \textbf{\bibinfo{volume}{121}}, \bibinfo{pages}{2813}
  (\bibinfo{year}{2004}).

\bibitem[{\citenamefont{Tarzia and Coniglio}(2006)}]{tarzia2006}
\bibinfo{author}{\bibfnamefont{M.}~\bibnamefont{Tarzia}} \bibnamefont{and}
  \bibinfo{author}{\bibfnamefont{A.}~\bibnamefont{Coniglio}},
  \bibinfo{journal}{Phys. Rev. Lett.} \textbf{\bibinfo{volume}{96}},
  \bibinfo{pages}{075702} (\bibinfo{year}{2006}).

\bibitem[{\citenamefont{Tarzia and Coniglio}(2007)}]{tarzia2007}
\bibinfo{author}{\bibfnamefont{M.}~\bibnamefont{Tarzia}} \bibnamefont{and}
  \bibinfo{author}{\bibfnamefont{A.}~\bibnamefont{Coniglio}},
  \bibinfo{journal}{Phys. Rev. E} \textbf{\bibinfo{volume}{75}},
  \bibinfo{pages}{011410} (\bibinfo{year}{2007}).

\bibitem[{\citenamefont{Archer and Wilding}(2007)}]{archer2007}
\bibinfo{author}{\bibfnamefont{A.~J.} \bibnamefont{Archer}} \bibnamefont{and}
  \bibinfo{author}{\bibfnamefont{N.~B.} \bibnamefont{Wilding}},
  \bibinfo{journal}{Phys. Rev. E} \textbf{\bibinfo{volume}{76}},
  \bibinfo{pages}{031501} (\bibinfo{year}{2007}).

\bibitem[{\citenamefont{Likos}(2001)}]{likos2001}
\bibinfo{author}{\bibfnamefont{C.}~\bibnamefont{Likos}},
  \bibinfo{journal}{Physics Reports} \textbf{\bibinfo{volume}{348}},
  \bibinfo{pages}{267} (\bibinfo{year}{2001}).

\bibitem[{\citenamefont{Royall and Williams}(2011)}]{royall2011c60}
\bibinfo{author}{\bibfnamefont{C.~P.} \bibnamefont{Royall}} \bibnamefont{and}
  \bibinfo{author}{\bibfnamefont{S.~R.} \bibnamefont{Williams}},
  \bibinfo{journal}{J. Phys. Chem. B} \textbf{\bibinfo{volume}{115}},
  \bibinfo{pages}{7288} (\bibinfo{year}{2011}).

\bibitem[{\citenamefont{Cates et~al.}(2004)\citenamefont{Cates, Fuchs, Kroy,
  and Puertas}}]{cates2004}
\bibinfo{author}{\bibfnamefont{M.}~\bibnamefont{Cates}},
  \bibinfo{author}{\bibfnamefont{M.}~\bibnamefont{Fuchs}},
  \bibinfo{author}{\bibfnamefont{W.~C.~K.} \bibnamefont{Kroy},
  \bibfnamefont{K.~Poon}}, \bibnamefont{and}
  \bibinfo{author}{\bibfnamefont{A.~M.} \bibnamefont{Puertas}},
  \bibinfo{journal}{J. Phys. Condens} \textbf{\bibinfo{volume}{16}},
  \bibinfo{pages}{S4861} (\bibinfo{year}{2004}).

\bibitem[{\citenamefont{Asakura and Oosawa}(1954)}]{asakura1954}
\bibinfo{author}{\bibfnamefont{S.}~\bibnamefont{Asakura}} \bibnamefont{and}
  \bibinfo{author}{\bibfnamefont{F.}~\bibnamefont{Oosawa}}, \bibinfo{journal}{J
  Chem Phys} \textbf{\bibinfo{volume}{22}}, \bibinfo{pages}{1255}
  (\bibinfo{year}{1954}).

\bibitem[{\citenamefont{Soga et~al.}(1998)\citenamefont{Soga, Melrose, and
  Ball}}]{soga1998}
\bibinfo{author}{\bibfnamefont{K.~G.} \bibnamefont{Soga}},
  \bibinfo{author}{\bibfnamefont{J.~R.} \bibnamefont{Melrose}},
  \bibnamefont{and} \bibinfo{author}{\bibfnamefont{R.~C.} \bibnamefont{Ball}},
  \bibinfo{journal}{J. Chem. Phys.} \textbf{\bibinfo{volume}{108}},
  \bibinfo{pages}{6026} (\bibinfo{year}{1998}).

\bibitem[{\citenamefont{d'Arjuzon et~al.}(2003)\citenamefont{d'Arjuzon, Frith,
  and Melrose}}]{darjuzon2003}
\bibinfo{author}{\bibfnamefont{R.~J.~M.} \bibnamefont{d'Arjuzon}},
  \bibinfo{author}{\bibfnamefont{W.}~\bibnamefont{Frith}}, \bibnamefont{and}
  \bibinfo{author}{\bibfnamefont{J.~R.} \bibnamefont{Melrose}},
  \bibinfo{journal}{Phys. Rev. E.} \textbf{\bibinfo{volume}{67}},
  \bibinfo{pages}{061404} (\bibinfo{year}{2003}).

\bibitem[{\citenamefont{Fortini et~al.}(2008)\citenamefont{Fortini, Sanz, and
  Dijkstra}}]{fortini2008}
\bibinfo{author}{\bibfnamefont{A.}~\bibnamefont{Fortini}},
  \bibinfo{author}{\bibfnamefont{E.}~\bibnamefont{Sanz}}, \bibnamefont{and}
  \bibinfo{author}{\bibfnamefont{M.}~\bibnamefont{Dijkstra}},
  \bibinfo{journal}{Phys. Rev. E} \textbf{\bibinfo{volume}{78}},
  \bibinfo{pages}{041402} (\bibinfo{year}{2008}).

\bibitem[{\citenamefont{Assoud et~al.}(2009)\citenamefont{Assoud, Ebert, Keim,
  Messina, Maret, and Loewen}}]{assoud2009}
\bibinfo{author}{\bibfnamefont{L.}~\bibnamefont{Assoud}},
  \bibinfo{author}{\bibfnamefont{F.}~\bibnamefont{Ebert}},
  \bibinfo{author}{\bibfnamefont{P.}~\bibnamefont{Keim}},
  \bibinfo{author}{\bibfnamefont{R.}~\bibnamefont{Messina}},
  \bibinfo{author}{\bibfnamefont{G.}~\bibnamefont{Maret}}, \bibnamefont{and}
  \bibinfo{author}{\bibfnamefont{H.}~\bibnamefont{Loewen}},
  \bibinfo{journal}{Phys. Rev. Lett.} \textbf{\bibinfo{volume}{102}},
  \bibinfo{pages}{238301} (\bibinfo{year}{2009}).

\bibitem[{\citenamefont{Alsayed et~al.}(2004)\citenamefont{Alsayed, Dogic, and
  Yodh}}]{alsayed2004}
\bibinfo{author}{\bibfnamefont{A.}~\bibnamefont{Alsayed}},
  \bibinfo{author}{\bibfnamefont{Z.}~\bibnamefont{Dogic}}, \bibnamefont{and}
  \bibinfo{author}{\bibfnamefont{A.}~\bibnamefont{Yodh}},
  \bibinfo{journal}{Phys. Rev. Lett.} \textbf{\bibinfo{volume}{93}},
  \bibinfo{pages}{057801} (\bibinfo{year}{2004}).

\bibitem[{\citenamefont{Taylor and Royall}(2012)}]{taylor2012}
\bibinfo{author}{\bibfnamefont{S.}~\bibnamefont{Taylor}} \bibnamefont{and}
  \bibinfo{author}{\bibfnamefont{C.~P.} \bibnamefont{Royall}},
  \bibinfo{journal}{Submitted}  (\bibinfo{year}{2012}).

\bibitem[{\citenamefont{Elsner et~al.}(2009)\citenamefont{Elsner, Snoswell,
  Royall, and Vincent}}]{elsner2009}
\bibinfo{author}{\bibfnamefont{N.}~\bibnamefont{Elsner}},
  \bibinfo{author}{\bibfnamefont{D.~R.~E.} \bibnamefont{Snoswell}},
  \bibinfo{author}{\bibfnamefont{C.~P.} \bibnamefont{Royall}},
  \bibnamefont{and} \bibinfo{author}{\bibfnamefont{B.~V.}
  \bibnamefont{Vincent}}, \bibinfo{journal}{J. Chem. Phys.}
  \textbf{\bibinfo{volume}{130}}, \bibinfo{pages}{154901}
  (\bibinfo{year}{2009}).

\bibitem[{\citenamefont{Hertlein et~al.}(2008)\citenamefont{Hertlein, Helden,
  Gambassi, Dietrich, and Bechinger}}]{hertlein2008}
\bibinfo{author}{\bibfnamefont{C.}~\bibnamefont{Hertlein}},
  \bibinfo{author}{\bibfnamefont{L.}~\bibnamefont{Helden}},
  \bibinfo{author}{\bibfnamefont{A.}~\bibnamefont{Gambassi}},
  \bibinfo{author}{\bibfnamefont{S.}~\bibnamefont{Dietrich}}, \bibnamefont{and}
  \bibinfo{author}{\bibfnamefont{C.}~\bibnamefont{Bechinger}},
  \bibinfo{journal}{Nature} \textbf{\bibinfo{volume}{172-175}},
  \bibinfo{pages}{451} (\bibinfo{year}{2008}).

\bibitem[{\citenamefont{Guo et~al.}(2007)\citenamefont{Guo, Narayanan, Sztuchi,
  Schall, and Wegdam}}]{guo2008}
\bibinfo{author}{\bibfnamefont{H.}~\bibnamefont{Guo}},
  \bibinfo{author}{\bibfnamefont{T.}~\bibnamefont{Narayanan}},
  \bibinfo{author}{\bibfnamefont{M.}~\bibnamefont{Sztuchi}},
  \bibinfo{author}{\bibfnamefont{P.}~\bibnamefont{Schall}}, \bibnamefont{and}
  \bibinfo{author}{\bibfnamefont{G.~H.} \bibnamefont{Wegdam}},
  \bibinfo{journal}{Phys. Rev. Lett.} \textbf{\bibinfo{volume}{100}},
  \bibinfo{pages}{188203} (\bibinfo{year}{2007}).

\bibitem[{\citenamefont{Bonn et~al.}(2009)\citenamefont{Bonn, Otwinowski,
  Sacanna, Guo, Wegdam, and Schall}}]{bonn2009}
\bibinfo{author}{\bibfnamefont{D.}~\bibnamefont{Bonn}},
  \bibinfo{author}{\bibfnamefont{J.}~\bibnamefont{Otwinowski}},
  \bibinfo{author}{\bibfnamefont{S.}~\bibnamefont{Sacanna}},
  \bibinfo{author}{\bibfnamefont{H.}~\bibnamefont{Guo}},
  \bibinfo{author}{\bibfnamefont{G.}~\bibnamefont{Wegdam}}, \bibnamefont{and}
  \bibinfo{author}{\bibfnamefont{P.}~\bibnamefont{Schall}},
  \bibinfo{journal}{Phy. Rev. Lett.} \textbf{\bibinfo{volume}{103}},
  \bibinfo{pages}{156101} (\bibinfo{year}{2009}).

\bibitem[{\citenamefont{Royall et~al.}(2008)\citenamefont{Royall, Williams,
  Ohtsuka, and Tanaka}}]{royall2008g}
\bibinfo{author}{\bibfnamefont{C.~P.} \bibnamefont{Royall}},
  \bibinfo{author}{\bibfnamefont{S.~R.} \bibnamefont{Williams}},
  \bibinfo{author}{\bibfnamefont{T.}~\bibnamefont{Ohtsuka}}, \bibnamefont{and}
  \bibinfo{author}{\bibfnamefont{H.}~\bibnamefont{Tanaka}},
  \bibinfo{journal}{Nature Mater.} \textbf{\bibinfo{volume}{7}},
  \bibinfo{pages}{556} (\bibinfo{year}{2008}).

\bibitem[{\citenamefont{Williams}(2007)}]{williams2007}
\bibinfo{author}{\bibfnamefont{S.~R.} \bibnamefont{Williams}},
  \bibinfo{journal}{Cond.Mat.ArXiV}
  \textbf{\bibinfo{volume}{ArXiv:0705.0203v1}} (\bibinfo{year}{2007}).

\bibitem[{\citenamefont{Royall et~al.}(2007)\citenamefont{Royall, Louis, and
  Tanaka}}]{royall2007}
\bibinfo{author}{\bibfnamefont{C.~P.} \bibnamefont{Royall}},
  \bibinfo{author}{\bibfnamefont{A.~A.} \bibnamefont{Louis}}, \bibnamefont{and}
  \bibinfo{author}{\bibfnamefont{H.}~\bibnamefont{Tanaka}},
  \bibinfo{journal}{J. Chem. Phys.} \textbf{\bibinfo{volume}{127}},
  \bibinfo{eid}{044507} (pages~\bibinfo{numpages}{8}) (\bibinfo{year}{2007}).

\bibitem[{\citenamefont{Noro and Frenkel}(2000)}]{noro2000}
\bibinfo{author}{\bibfnamefont{M.~G.} \bibnamefont{Noro}} \bibnamefont{and}
  \bibinfo{author}{\bibfnamefont{D.}~\bibnamefont{Frenkel}},
  \bibinfo{journal}{J. Chem. Phys.} \textbf{\bibinfo{volume}{113}},
  \bibinfo{pages}{2941} (\bibinfo{year}{2000}).

\bibitem[{\citenamefont{Elliot and Hu}(1999)}]{eliott1999}
\bibinfo{author}{\bibfnamefont{J.~R.} \bibnamefont{Elliot}} \bibnamefont{and}
  \bibinfo{author}{\bibfnamefont{L.}~\bibnamefont{Hu}}, \bibinfo{journal}{J.
  Chem. Phys.} \textbf{\bibinfo{volume}{110}}, \bibinfo{pages}{3043}
  (\bibinfo{year}{1999}).

\bibitem[{\citenamefont{Doye et~al.}(1995)\citenamefont{Doye, Wales, and
  Berry}}]{doye1995}
\bibinfo{author}{\bibfnamefont{J.~P.~K.} \bibnamefont{Doye}},
  \bibinfo{author}{\bibfnamefont{D.~J.} \bibnamefont{Wales}}, \bibnamefont{and}
  \bibinfo{author}{\bibfnamefont{R.~S.} \bibnamefont{Berry}},
  \bibinfo{journal}{J. Chem. Phys.} \textbf{\bibinfo{volume}{103}},
  \bibinfo{pages}{4234} (\bibinfo{year}{1995}).

\bibitem[{\citenamefont{Malins et~al.}(2012)\citenamefont{Malins, Eggers, and
  Royall}}]{malins2012}
\bibinfo{author}{\bibfnamefont{A.}~\bibnamefont{Malins}},
  \bibinfo{author}{\bibfnamefont{J.}~\bibnamefont{Eggers}}, \bibnamefont{and}
  \bibinfo{author}{\bibfnamefont{C.~P.} \bibnamefont{Royall}},
  \bibinfo{journal}{Submitted}  (\bibinfo{year}{2012}).

\bibitem[{\citenamefont{Frank}(1952)}]{frank1952}
\bibinfo{author}{\bibfnamefont{F.~C.} \bibnamefont{Frank}},
  \bibinfo{journal}{Proc. R. Soc. Lond. A.} \textbf{\bibinfo{volume}{215}},
  \bibinfo{pages}{43} (\bibinfo{year}{1952}).

\bibitem[{\citenamefont{Taffs et~al.}(2010)\citenamefont{Taffs, Malins,
  Williams, and Royall}}]{taffs2010jcp}
\bibinfo{author}{\bibfnamefont{J.}~\bibnamefont{Taffs}},
  \bibinfo{author}{\bibfnamefont{A.}~\bibnamefont{Malins}},
  \bibinfo{author}{\bibfnamefont{S.~R.} \bibnamefont{Williams}},
  \bibnamefont{and} \bibinfo{author}{\bibfnamefont{C.~P.}
  \bibnamefont{Royall}}, \bibinfo{journal}{J. Chem. Phys.}
  \textbf{\bibinfo{volume}{133}}, \bibinfo{pages}{244901}
  (\bibinfo{year}{2010}).

\bibitem[{\citenamefont{Malins et~al.}(2009)\citenamefont{Malins, Williams,
  Eggers, Tanaka, and Royall}}]{malins2009}
\bibinfo{author}{\bibfnamefont{A.}~\bibnamefont{Malins}},
  \bibinfo{author}{\bibfnamefont{S.~R.} \bibnamefont{Williams}},
  \bibinfo{author}{\bibfnamefont{J.}~\bibnamefont{Eggers}},
  \bibinfo{author}{\bibfnamefont{H.}~\bibnamefont{Tanaka}}, \bibnamefont{and}
  \bibinfo{author}{\bibfnamefont{C.~P.} \bibnamefont{Royall}},
  \bibinfo{journal}{J. Phys.: Condens. Matter} \textbf{\bibinfo{volume}{21}},
  \bibinfo{pages}{425103} (\bibinfo{year}{2009}).

\bibitem[{\citenamefont{Zhang et~al.}(2012{\natexlab{b}})\citenamefont{Zhang,
  Klok, Hans~Tromp, Groenewold, and Kegel}}]{zhang2011cluster}
\bibinfo{author}{\bibfnamefont{T.~H.} \bibnamefont{Zhang}},
  \bibinfo{author}{\bibfnamefont{J.}~\bibnamefont{Klok}},
  \bibinfo{author}{\bibfnamefont{R.}~\bibnamefont{Hans~Tromp}},
  \bibinfo{author}{\bibfnamefont{J.}~\bibnamefont{Groenewold}},
  \bibnamefont{and} \bibinfo{author}{\bibfnamefont{W.~K.} \bibnamefont{Kegel}},
  \bibinfo{journal}{Soft Matter} \textbf{\bibinfo{volume}{8}},
  \bibinfo{pages}{667} (\bibinfo{year}{2012}{\natexlab{b}}).

\bibitem[{\citenamefont{Wilding and Sollich}(2011)}]{wilding2011}
\bibinfo{author}{\bibfnamefont{N.~B.} \bibnamefont{Wilding}} \bibnamefont{and}
  \bibinfo{author}{\bibfnamefont{P.}~\bibnamefont{Sollich}},
  \bibinfo{journal}{Soft Matter} \textbf{\bibinfo{volume}{7}},
  \bibinfo{pages}{4472} (\bibinfo{year}{2011}).

\bibitem[{\citenamefont{Klotsa and Jack}(2011)}]{klotsa2011}
\bibinfo{author}{\bibfnamefont{D.}~\bibnamefont{Klotsa}} \bibnamefont{and}
  \bibinfo{author}{\bibfnamefont{R.~L.} \bibnamefont{Jack}},
  \bibinfo{journal}{Soft Matter} \textbf{\bibinfo{volume}{7}},
  \bibinfo{pages}{6294} (\bibinfo{year}{2011}).

\bibitem[{\citenamefont{Foffi et~al.}(2005)\citenamefont{Foffi, De~Michele,
  Sciortino, and Tartaglia}}]{foffi2005jcp}
\bibinfo{author}{\bibfnamefont{G.}~\bibnamefont{Foffi}},
  \bibinfo{author}{\bibfnamefont{C.}~\bibnamefont{De~Michele}},
  \bibinfo{author}{\bibfnamefont{F.}~\bibnamefont{Sciortino}},
  \bibnamefont{and}
  \bibinfo{author}{\bibfnamefont{P.}~\bibnamefont{Tartaglia}},
  \bibinfo{journal}{J. Chem. Phys} \textbf{\bibinfo{volume}{122}},
  \bibinfo{pages}{224903} (\bibinfo{year}{2005}).

\bibitem[{\citenamefont{Furukawa and Tanaka}(2010)}]{furukawa2010}
\bibinfo{author}{\bibfnamefont{A.}~\bibnamefont{Furukawa}} \bibnamefont{and}
  \bibinfo{author}{\bibfnamefont{H.}~\bibnamefont{Tanaka}},
  \bibinfo{journal}{Phys. Rev. Lett.} \textbf{\bibinfo{volume}{104}},
  \bibinfo{pages}{245702} (\bibinfo{year}{2010}).

\end{thebibliography}

\end{document}